\documentclass[debug]{rmaa}

\usepackage{mathtools}
\usepackage[T1]{fontenc}
\usepackage[para,online,flushleft]{threeparttable}
\usepackage{paralist}
\usepackage[nottoc]{tocbibind}

\usepackage{psfrag,color}
\usepackage{lineno,hyperref}
\usepackage{comment}
\usepackage{euscript}
\usepackage{array}
\usepackage{paralist}
\usepackage{subfigure}
\usepackage{amsmath}
\usepackage{color}
\usepackage{amssymb}
\usepackage[latin1]{inputenc}
\usepackage{float}
\usepackage[normalem]{ulem}

\usepackage{tabto}
\usepackage{diagbox}
\usepackage{rotating,booktabs}
\usepackage{euscript}
\usepackage{lineno}

\title{Optimal transfers from Moon to $L_2$ halo orbit of the Earth-Moon system.} 

\author{
 Santos, L. B. T.;\altaffilmark{1} 
  de Almeida Jr, Allan Kardec;\altaffilmark{1,2}
  Sousa-Silva, P. A.;\altaffilmark{3} 
  Terra, M. O.;\altaffilmark{4} 
  Sanchez, D. M.;\altaffilmark{5} 
  Aljbaae S. ;\altaffilmark{1} 
  Prado, A. F. B. A.;\altaffilmark{6,7} 
  and Monteiro, F.\altaffilmark{8}}

\altaffiltext{1}{Division of Space Mechanics and Control. National Institute for Space Research, INPE. S\~ao Jos\'e dos Campos, Brazil.}
\altaffiltext{2}{Instituto de Telecomunica\c{c}\~oes, 3810-193 Aveiro, Portugal (allan.junior@inpe.br).}
\altaffiltext{3}{S\~ao Paulo State University, UNESP. S\~ao Jo\~ao da Boa Vista, Brazil. (priscilla.silva@unesp.br).}
\altaffiltext{4}{Technological Institute of Aeronautics, ITA. S\~ao Paulo dos Campos, Brazil. (maisa@ita.br).}
\altaffiltext{5}{The University of Oklahoma, Norman OK, USA.}
\altaffiltext{6}{Postgraduate Division - National Institute for Space Research (INPE), S\~ao Paulo, Brazil.} 
\altaffiltext{7}{Professor, Academy of Engineering, RUDN University, Miklukho-Maklaya street 6, Moscow, Russia, 117198. (antonio.prado@inpe.br).}
\altaffiltext{8}{National Observatory, ON. Rio de Janeiro, Brazil. (filipeastro@on.br).}

\shortauthor{Santos, L. B. T., et al}
\shorttitle{ORBITAL DYNAMICS AROUND A BINARY ASTEROID SYSTEM.}

\listofauthors{L. B. T. dos Santos, de Almeida Jr, Allan Kardec, P. A. Sousa-Silva, Terra, M. O., D. M. Sanchez, Safwan, A.  \& A. F. B. A. Prado}
\indexauthor{Santos, L. B. T.}
\indexauthor{de Almeida Jr, Allan Kardec}
\indexauthor{Sousa-Silva, P. A.}
\indexauthor{Terra, M. O.}
\indexauthor{Sanchez, D. M.}
\indexauthor{Aljbaae, S.}
\indexauthor{Prado, A. F. B. A.}

\abstract{
In this paper, we seek optimal solutions for a transfer from a parking orbit around the Moon to a halo orbit around $L_2$ of the Earth-Moon system, by applying a single maneuver and exploiting the stable invariant manifold of the hyperbolic parking solution at arrival.
For that, we propose an optimization problem considering as variables both the orbital characteristics of a parking solution around the Moon, (namely, its Keplerian elements) and the characteristics of a transfer trajectory guided by the stable manifold of the arrival Halo orbit. The problem is solved by a nonlinear programming method (NLP), aiming to minimize the cost of $\Delta V$ to perform a single maneuver transfer, within the framework of the Earth-Moon system of the circular restricted three-body problem. Results with low $\Delta V$ and suitable time of flight show the feasibility of this kind of transfer for a Cubesat.}

\addkeyword{Stability}
\addkeyword{Celestial mechanics}
\addkeyword{Minor planets, asteroids: general}

\begin{document}
\maketitle

\section{Introduction}
\label{intro}
The Apollo Program has brought significant gains to humanity in science and technology, after all, it is not surprising that in the 1970s the first microcomputers emerged. Owing to advances in technology miniaturization, a new class of small spacecraft, known in the space industry as CubeSat, have been built. Compared to traditional satellites, a CubeSat greatly reduces the complexity, and costs of development and launch. Owing to these factors, a dramatic increase in CubeSat launches has been observed over the last decade \citep{Poghosyan}. Since they are light and small, Cubesats can be launched as a secondary payload by any rocket that sends a conventional satellite into space and positioned in many classes of orbits. From these orbits, the Cubesats can be transferred to their operational orbits to perform their observation/science missions. 

Our motivation is to address the need for optimized trajectories since there a growing interest in missions using a Cubesat aimed at exploring the Cislunar environment. An example is Exploration Mission 1 which selected 13 CubeSats to use as secondary payloads. In 2021, the NASA plans to launch a Cubesat (mission Lunar Flashlight) to explore the Moon for the purpose of finding water ice hidden in areas that are permanently shaded, which may be used for human exploration \citep{Peter}. Other missions, such as OMOTENASHI \citep{Campagnola} and Lunar Lantern \citep{Lai}, are designed to fulfill a wide range of scientific and technological objectives using a semi-hard landing on the lunar surface and observing water on the Moon, respectively \citep{Chen}. EQUULEUS aims to reach the equilibrium point $L_2$ of the Earth-Moon system to analyze the far side of the Moon. The halo orbit around $L_2$ of the Earth-Moon system maintains the geometrical configuration and constant distances from the Earth and the Moon. These characteristics favor Cubesat thermal control, communication between the dark side of the Moon and Earth, among others \citep{Far, Chen}. 

As shown in the \citet{Parker} monograph in Sec. 3.5.1, there is a wide range of trajectories that fly by with different radii (distance from the center of the Moon) and inclination around the Moon that can be reached naturally through an unstable manifold, starting from halo orbits of the $L_2$ family.

\citet{Trofimov} conducted a study using an impulsive transfer maneuver directed to the Moon from a Near-rectilinear halo orbit to the intermediate low-perilune orbit (LPO). The author found results that show that most of the optimal solutions reach the Moon with an inclination between 60$^\circ$ and 120$^\circ$. The $\Delta$V values of the optimal solutions oscillate around 100 and 200 m/s, depending on the inclination and the distance from the Moon that the trajectory reaches.

The scientific community is now looking for trajectories in the vicinity of the Moon. Because CubeSat has a small size, the thrust has limitations due to low mass propellant, these trajectories should be of low energy, utilizing low transfer time and affordable thruster requirements \citep{Alessi}. 

Keeping this in mind, this article concerns the mission design from a lunar parking orbit to a parking halo orbit of the central manifold of the $L_2$ equilibrium point, with a low cost and adequate time of flight. For that, we consider an optimization problem in which we vary at the same time the orbital properties of the departure, transfer and arrival orbits, in order to generate such an optimal solution.

To fulfill the objectives of this paper, a halo orbit around the $L_2$ equilibrium point of the Earth-Moon system is used as a reference orbit. So the question that arises is: ``What is the most efficient and economical way to reach halo orbit from a parking orbit around the Moon?''

Knowing that the orbits around the collinear equilibrium points have a strong hyperbolic characteristic, as stated in \citet{Gomez2, Gomez3}, this feature can be used to perform a transfer using less fuel ($\Delta V$) and reach the desired halo (or Lyapunov) orbit from a parking orbit. This hyperbolic trajectory that asymptotically departs or arrives in a periodic orbit around an equilibrium point is called the unstable or stable invariant manifold, respectively. 

\subsection{Propulsion for a Stand-Alone Cislunar environment CubeSat}

Propulsion system becomes indispensable for stand-alone CubeSats bound for the Cislunar environment due to the need for complex orbit maneuvering and precise trajectory control. 

There are several types of propulsion that are applicable for stand-alone deep-space CubeSats \citep{mani2020combined}. The choice of the propulsion system depends upon the mission objectives and is at the discretion of the spacecraft designer. Chemical propulsion systems applicable for CubeSats include liquid monopropellant systems, liquid bi-propellant systems, and miniaturized solid rocket motors. Electric propulsion systems applicable for CubeSats include miniaturized gridded ion thrusters, Hall thrusters, Field Emission Electric Propulsion etc. Additionally, a third category called multimode propulsion is also applicable for CubeSats \citep{rovey2020review}.

The transfer to the Cislunar environment can be achieved using fully chemical, fully electric, or combined chemical-electric propulsion solutions. Fully chemical transfers lead to shorter transfer times but larger system mass due to high thrust low specific impulse of chemical propulsion systems. Fully electric transfers lead to lower system mass but longer transfer times due to high specific impulse low thrust performance of electric propulsion systems. Combined chemical--electric propulsion balances the transfer time and system mass \citep{mani2018combined}. The mission needs play a key role since several factors influence the design. This include overall mass requirement, cost of spacecraft operations, precision control and pointing requirements, and requirement on the timeline for commencement of operations. 

Comparison of Chemical Propulsion Systems and Electric Propulsion Systems are shown in Table.~\ref{tab:chemproptradeoff} \citep{gohardani2014green, lemmer2017propulsion} and Table.~\ref{tab:elecpropcomparison} \citep{mazouffre2016electric}, respectively.

\begin{sidewaystable}
\centering
\caption{Comparison of chemical propulsion systems}
\label{tab:chemproptradeoff}
\begin{threeparttable}
\begin{tabular}{c c c c c c c c}
\toprule
\toprule
\textbf{Type} & \textbf{Thrust level} & $\mathbf{I_{sp}}$ & \textbf{Mass} & \textbf{Complexity} & \textbf{Control} & \textbf{Reliability} & \textbf{TRL} \\
\hline
Cold gas & \footnotesize Very low, $<$1 N & \footnotesize Very low, $<$80 s & \footnotesize Very large for req. $\Delta V$ & \footnotesize Simple & \footnotesize Short burn duration & \footnotesize High & 9 \\ 

Solid & \footnotesize Very high, $>$5 N & \footnotesize Moderate 200--250 s & \footnotesize Large for req. $\Delta V$ & \footnotesize Simple & \footnotesize No start-stop capability & \footnotesize Low & 9 \\ 

Monopropellant & \footnotesize Moderate, 1--1.5 N & \footnotesize High 230--270 s & \footnotesize Moderate for req. $\Delta V$ & \footnotesize Relatively complicated & \footnotesize Feed system needs minor effort & \footnotesize Good & 7 \\ 

Bipropropellant & \footnotesize High 1.5--1.9 N & \footnotesize Very high 300--330 s & \footnotesize Low for req. $\Delta V$ & \footnotesize Extremely complicated & \footnotesize Feed system requires major effort & \footnotesize Very Low & 4 \\ 
\bottomrule
\bottomrule
\end{tabular}
\end{threeparttable}
\end{sidewaystable}

\begin{sidewaystable}
\centering
\caption{Comparison of electric propulsion options}
\label{tab:elecpropcomparison}
\begin{tabular}{c c c c c}
\toprule
\toprule
{\textbf{Type}} & \textbf{Thrust [mN]} & \textbf{$\mathbf{I_{sp}}$ [s]} & \textbf{Power [W]} & \textbf{Life [\textcolor{black}{hours}]}\\
\hline
Gridded Ion & 0.5-1.6 	& 1500-3200 &  30-80  & $\sim$30000 \\ 
Hall & 1.8-4  	& 800-1400  &  60-120 & $\sim$10000  \\
FEEP	&	0.35-2 &	$>$6000 & 28-160	&	$<$10000 \\
Pulsed Plasma & 0.01-1 	& 500-1500	&  10-30  & $\sim$1000 \\
Helicon & 0.8-1.5	&	900-1200 & 50-80	& $\sim$1000 \\
\bottomrule
\bottomrule
\end{tabular}%
\end{sidewaystable}
Selection of the propellant system is done based on a reference CubeSat mission of 12U volume and 24 kg wet mass requiring a $\Delta V$ of 200 m/s.

From Table.~\ref{tab:chemproptradeoff}, it can be observed that the monopropellant system is the best suited chemical propulsion system in terms of a trade-off between thrust, $I_{sp}$, mass, complexity, control, reliability and technology readiness level (TRL). For a monopropellant system an average $I_{sp}$ of 250 s, the required propellant mass is approximately 1.9 kg. Although Bipropellant systems would lead to a lower propellant mass, the corresponding feed system design and integration is complicated for CubeSat applications. Cold gas systems yield a much larger propellant mass and the solid rocket motors are not controllable in orbit.

Regarding electric propulsion systems, given that the $I_{sp}$ is very high and the thrust is very low, the required propellant mass will be very low but the transfer time will be very high. From Table.~\ref{tab:elecpropcomparison}, considering gridded ion thrusters with an average $I_{sp}$ of 2200 s, the propellant mass required is only 0.25 kg. However, one has to take into account the added mass of the solar arrays and power systems that are needed to supply the required power to the solar electric propulsion systems.

\subsection{Statement of the problem}

In this article, we analyze impulsive maneuvers to perform a transfer from an orbit around the Moon to the halo orbit around $L_2$ equilibrium point of the Earth-Moon system. For this work, we employ an invariant manifold to reach halo orbits. The design transfers are defined as follows: Cubesat is initially in a parking orbit around the Moon. From this, we need to position the Cubesat tangent to the stable manifold  as precisely as possible. This stable invariant manifold will asymptotically reach the halo orbit around the equilibrium point $L_2$ of the Earth-Moon system. In detail, we seek the intersection in the configuration space between the parking orbit around the Moon (selenocentric orbit) and the stable invariant manifold. 

The objective of the transfer design analysis is to find the optimal orbital characteristics (Keplerian elements) of a particle around the Moon. Optimal orbital characteristics are those that minimize total fuel consumption ($\Delta V$) to perform transfers between the selenocentric orbit and the stable invariant manifold. The optimization problem is defined and solved with a nonlinear programming method (NLP). Regarding the orbital maneuvers performed by a Cubesat, we look for solutions whose $\Delta V$ $\leq$ 200 m/s. The maneuver that inserts the spacecraft into the stable invariant manifold and connects the target halo is called the Insertion Maneuver ($IM$), $\Delta V$ \citep{Dio}. 

\section{Equations of motion}

In this model, the motion of a negligible mass ($P(x,~y,~z)$) is governed by the gravitational forces of the primary bodies $P_1$ and $P_2$, respectively, with mass $m_1$ and $m_2$.  It is assumed that the infinitesimal mass body does not affect the dynamics of the primaries, which perform circular orbits around their barycenter. Distances are normalized by the distance between $P_1$ and $P_2$, while the unit of time is defined such that the orbital period of $P_1$ and $P_2$ around the center of mass is equal to $2\pi$.  Given that $m_1 > m_2$, the mass ratio is defined as $\mu = m_2/(m_1+m_2)$.  The primary bodies $P_1$ and $P_2$ (of normalized mass $1-\mu$ and $\mu$, respectively) are on the $x$ axis, with respect to the rotating frame, whose positions are $x_1$ = $-\mu$ and $x_2$ = $1-\mu$, respectively. 

Let $\textit{\textbf{x}}$ = $[\boldsymbol{\rho}, \boldsymbol{\eta}]^T$ the state vector, in time t, of the particle $P$ in the restricted problem where the vector $\textit{\textbf{x}}$ $\in$  $\mathbb{R}^6$ and $\boldsymbol{\rho}$ (position) and $\boldsymbol{\eta}$ (velocity) $\in$  $\mathbb{R}^3$. The equations of motion of the Circular Restricted Three-Body Problem (CRTBP) can be rewritten by six first-order differential equations as
\begin{equation}
\textit{\textbf{x$'$}} = {\textit{\textbf{f}}(\textit{\textbf{x}},t)},  
\label{x'}
\end{equation}
where $\textit{\textbf{f}}$ is given by
\begin{equation}
\textit{\textbf{f}}=
\begin{bmatrix}
f_1\\ 
f_2\\ 
f_3\\ 
f_4\\ 
f_5\\ 
f_6
\end{bmatrix}
= \begin{bmatrix}
\dot{x}\\ 
\dot{y}\\ 
\dot{z}\\ 
x-\frac{(1-\mu)(x+\mu)}{r_1^3}-\frac{\mu(x+\mu-1)}{r_2^3}+2\dot{y}\\ 
y-\frac{(1-\mu)y}{r_1^3}-\frac{\mu y}{r_2^3}-2\dot{x}\\ 
-\frac{(1-\mu)z}{r_1^3}-\frac{\mu z}{r_2^3},
\end{bmatrix} 
\label{f}
\end{equation}

$\mathit{\mathbf{x}}$

and $r_1$ and $r_2$ are the distances from P to $P_1$ and from P to $P_2$, given by $r_1= \sqrt{(x+\mu)^2+y^2+z^2}$ and $r_2= \sqrt{(x+\mu-1)^2+y^2+z^2}$, respectively.
The CRTBP has five static solutions, $\mathit{\mathbf{x}}_k$, namely, such that $\mathit{\mathbf{f}}~(\mathit{\mathbf{x}}_k, \mu)=0$, $k$ = 1,..,5. These are known as equilibrium points or lagrange points and are labeled as $L_k$. The collinear points ($L_1$, $L_2$, $L_3$) are located on the $x$-axis and the triangular points ($L_4$, $L_5$) are in the $xy$ plane. So, all equilibrium points there exist in the plane $z = 0$. 
Linearly, the collinear points behave as the product saddle $\times$ center $\times$ center \citep{Mingotti2}. 
The two centers are related to in-plane and out-of-plane motion and, therefore indicate that there are families of planar and vertical periodic orbits known as Lyapunov's orbits, \citep{Jorba1, Gomez3, Topputo16}, as well as quasi-periodic motions known as Lissajous orbits \citep{Howell88}.

When the frequencies in-plane and out-of-plane match, a special family of three-dimensional periodic orbits emerges, these orbits are known as halo orbit \citep{Howell84}. The factor saddle implies the existence of a two-dimensional unstable (stable) invariant manifold that departs (reach) from (the) periodic orbits \citep{Howell84, Howell88, Gomez2, Topputo16, Priscilla}.

The CRTBP admits an integral of motion, known as Jacobi integral, given by \citep{1967torp.book.....S, Topputo15}
\begin{equation}
J(\textit{\textbf{x}}, \mu) = \frac{2(1-\mu)}{r_1}+\frac{2\mu}{r_2}+x^2+y^2-(\dot{x}^2+\dot{y}^2+\dot{z}^2)+\mu(1-\mu),  
\label{J}
\end{equation}
which defines a five-dimensional manifold for a given energy level $C$
\begin{equation}
J(C, \mu)=\left\{\textit{\textbf{x}} \in \mathbb{R}^6| J(\textit{\textbf{x}}, \mu) =C\right\}.  
\label{JC}
\end{equation}

\subsection{Invariant Manifolds for periodic orbit }
\label{sec:5}

\begin{figure}[!htbp]
\centering\includegraphics[scale=0.6]{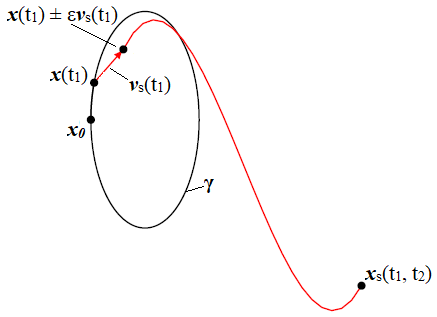}
\caption{Method used to compute the invariant manifolds.}
\label{Manifold1}
\end{figure}

Halo and Lyapunov orbits are two types of periodic solutions of the CRTBP that produce an interesting structure in the phase space. The phase space geometry investigated in this paper considers the first type of particular solution for the nonlinear differential equations and also considers the structure associated with local flow. A formidable reference for halo design and theory is \citep{Gomez1}. Once a Halo orbit is determined, we can discretize it into a large number of points, that can be parameterized by $t_1$ (parameter along the orbit), for $t_1 \in [0,T)$, with $T$ being the period of the halo orbit (as shown in Fig. \ref{Manifold1}). Thus, the trajectory of an invariant manifold reaching (or departing from) each point along the halo orbit can be computed. A manifold is a mathematical term used to refer to higher dimensional surfaces \citep{Koon}. In the theory of dynamical systems, an invariant manifold is a type of special manifold consisting of orbits. Any point located in an invariant manifold will remain forever in the manifold under the domain of the equations of motion. Invariant manifolds can be determined locally by the linear approximation of the dynamics. In this context, each point of the periodic orbit can be considered as a fixed point of a Time-T map, whose linear approximation is given by  the  monodromy matrix $M$.  By definition, $M$ is the State Transition Matrix (STM) of Equation \ref{f} propagated for the period $T$ of the orbit \citep{Natasha}. The STM of the Equation \ref{f}, given by $\boldsymbol{\Phi}(\textit{\textbf{x}}, t)=d\boldsymbol{\varphi}(\textit{\textbf{x}}, t)/d\textit{\textbf{x}}$, which is determined from the variational equation
\begin{equation}
\dot{\boldsymbol{\Phi}}(\textit{\textbf{x}}, t)=\frac{\mathrm{\partial}\textit{\textbf{f}}(\boldsymbol{\varphi}(\textit{\textbf{x}}(t),t),\mu)}{\mathrm{\partial}\textit{\textbf{x}}}\boldsymbol{\Phi}(\textit{\textbf{x}},t),~~~~ \boldsymbol{\Phi}(\textit{\textbf{x}},0) = I_{6x6}.
\label{Variational}
\end{equation}

Thus, the time evolution of $\boldsymbol{\Phi}(\textit{\textbf{x}},t)$ corresponds to the integration of 36 first-order differential equations, that must be performed simultaneously with the 6 first-order equations \ref{x'} and \ref{f}.

Given the symplectic nature of the monodromy matrix $M$, its eigenvalues are always given by three reciprocal pairs. Since one pair has unit value (orbit periodicity condition), two pairs remain to provide orbit stability information.
From Equation $\boldsymbol{M\nu_i}=\lambda_i\boldsymbol{\nu_i}$, for i=1,...,3 we can observe that the eigenvalues with real part smaller than one imply in  hyperbolic motion converging to the limit set, and  therefore, associated with the stable subspace ($\lambda_s$). 
So, in short, given the monodromy matrix $M(t_1)$, with $t_1$ $\in$ [0, $T$] the stable eigendirections are provided by the eigenvectors, $\boldsymbol{\nu_s}$($t_1$), that are associated with the eigenvalues $\lambda_s$($t_1$) $< 1$. These eigendirections provide the initial guesses for constructing the invariant manifolds of this halo orbit \citep{Howell88}.

Let $ \boldsymbol{\varphi}(\textit{\textbf{x}}_i, t)$ be the flow of Equation \ref{f} given by 
\begin{equation}
\boldsymbol{\varphi}(\textit{\textbf{x}}, t)=\textit{\textbf{x}}+\int_{0}^{t}\textit{\textbf{f}}(\textit{\textbf{x}}(\tau), \mu)d\tau.  
\label{Flow}
\end{equation}
Because CRTBP is autonomous, we can, without loss of generality, declare the initial time as zero. The sign of t (positive/negative) determine the direction of integration (forward/backward) \citep{Topputo16}.
Let $\boldsymbol{\gamma }$ be a generic periodic orbit in the CRTBP, as shown in Figure \ref{Manifold1}.
\begin{equation}
\boldsymbol{\gamma} =  \left\{\boldsymbol{\varphi}(\textit{\textbf{x}}_0, t)~|~t \in \mathbb{R}\right\},~~~~~~ \boldsymbol{\varphi}(\textit{\textbf{x}}_0, T) = \textit{\textbf{x}}_0,
\label{floww}
\end{equation}
where $\textit{\textbf{x}}_0$ is the initial state.

Based on Fig. \ref{Manifold1}, note that any point of $\boldsymbol{\gamma}$ can be determined as $\textit{\textbf{x}}(t_1) = \boldsymbol{\varphi}(\textbf{$x_0$},t_1)$. Suppose that $W^s(\boldsymbol{\gamma})$ is the stable two-dimensional manifolds of $\boldsymbol{\gamma}$. The main idea is to determine the one-dimensional manifolds of $\textit{\textbf{x}}(t_1)$ and, varying $t_1$, get $W^s(\boldsymbol{\gamma})$.

The stable eigenspace, $E_s$, is associated to the eigenvectors corresponding to the $\lambda_s<1$, given by $\boldsymbol{\nu}_s$. The $\boldsymbol{\nu}_s$ should be normalized \citep{Topputo16}. Tangent to this eigenspace there exist local stable manifolds, $x_{loc}^{s\pm}$, at the fixed point and with the same dimension \citep{Guckenheimer}.
The initial state vectors for determining local stable manifolds, can be calculated using
\begin{equation}
\textit{\textbf{x}}_{loc}^{s\pm} = \textit{\textbf{x}}(t_1) \pm \epsilon\boldsymbol{\nu}_s(t_1)  
\label{manifolds}
\end{equation}
where $\textit{\textbf{x}}_{loc}^{s\pm}$ $\in$ $W_{loc}^{s\pm}$ are generic points in the stable local manifold. $\epsilon$ is a small disturbance in the displacement in the same direction as the eigenvector. This little shift $\epsilon$ disturb $\textit{\textbf{x}}(t_1)$ towards the stable or unstable manifold, while the $\pm$ characterizes which of the two branches (inner or outer) of the manifold will be generated. Based on the arguments of \citet{Topputo16}, the value $\epsilon = 10^{-6}$ was used in this work. $\textit{\textbf{x}}_{loc}^{s\pm}$ produce trajectories that are in the vicinity of the periodic orbit $\boldsymbol{\gamma}$. To globalize manifold trajectories, state vectors ($\textit{\textbf{x}}_{loc}^{s\pm}$ must be propagated using the equations of motion (\ref{x'} and \ref{f}) of restricted problem (nonlinear equations). 

To globalize the stable invariant manifolds without any loss of generality, we define $\textit{\textbf{x}}_{s} \in W_s(\boldsymbol{\gamma})$ a generic point in the stable manifold, respectively, and we do
\begin{equation}
\textit{\textbf{x}}_{s}(t_1, t_2) = \boldsymbol{\varphi}(\textit{\textbf{x}}(t_1)\pm \epsilon\boldsymbol{\nu}_s(t_1), -t_2),
\label{unst}
\end{equation}
where $t_2$ (parameter along the flow) is the integration time, and $t_2>0$, $t_2$ $\in$  [0, $T_2$ ]; $T_2$ is an upper limit, depending on the extent to which $W_s(\boldsymbol{\gamma})$ must be globalized (see Figure \ref{Manifold1}).

\section{Methodology}

\begin{figure}
	\subfigure[\label{manifoldxy}]
	{
		\includegraphics[scale=0.35]{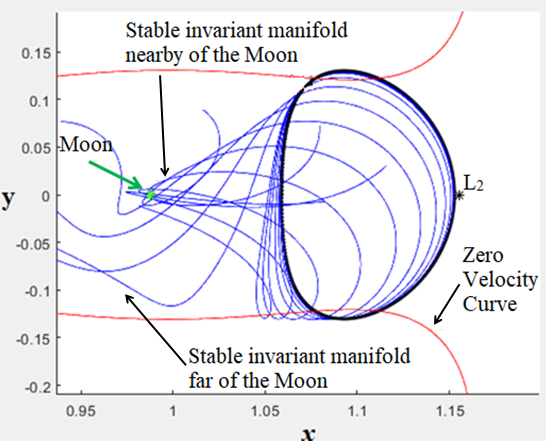}}
~
~
	\subfigure[\label{manifoldxyz}]
	{
		\includegraphics[scale=0.39]{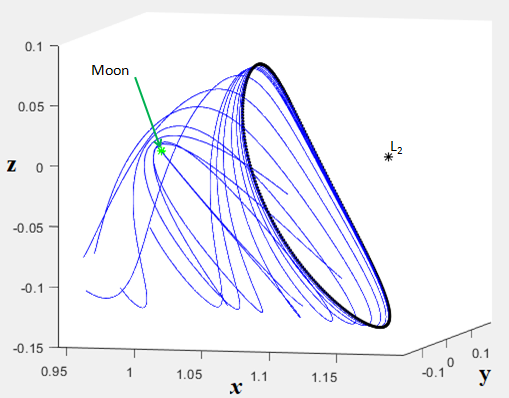}}
\caption{Stable invariant manifold trajectories of a halo orbit around $L_2$ of the Earth-Moon system projected (a) in the $xy$ plane, and (b) in the $xyz$ space.}
	\label{6Mmanifold}
\end{figure}

Figure \ref{6Mmanifold} presents 10 trajectories (blue curves) on the stable invariant manifold $W_s$ of a halo orbit around $L_2$ of the Earth-Moon system, illustrating a few of the infinite number of solutions contained in this invariant set.

It is noticeable from Figure \ref{6Mmanifold} that some solutions move closer to the Moon, while others move away. In this study, we are interested in solutions on the invariant manifolds that approach the Moon. This is due to the fact that, initially, the spacecraft (possibly a cubesat) will be in a parking orbit around the Moon. Thus, to perform the impulsive transfer maneuver minimizing fuel consumption ($\Delta V$), it is necessary to find the intersection between a solution of the stable invariant manifold of such halo orbit and the parking orbit.

For the first analysis, some restrictions of the Keplerian elements in the parking orbit will be adopted in this paper. In the parking orbit, we search for orbits whose altitudes of periselene and aposelene are 600 km and 20,000 km, respectively.

To find the optimal solutions, a nonlinear programming method (NLP) is applied. So, the transfer design problem with an impulsive maneuver aiming to minimize the $\Delta V$ can be stated as follows:

\textbf{Variables.} The NLP variables are the Keplerian elements of the parking orbit around the Moon and the computed times to determine stable manifolds. These variables are written in a vector $\textit{\textbf{y}}$, as shown in Equation \ref{y}
\begin{equation}
\textit{\textbf{y}} = (h_p,h_a, i,\Omega,\omega,\theta,t_1,t_2), 
\label{y}
\end{equation}
where $t_1$ determines a state $\textit{\textbf{x}}(t_1)$ along the halo orbit, and $t_2$ is the integration time from the parking orbit until the halo orbit. We remark that $t_2$ must be negative because we need to integrate backwards in time to find the stable manifold of the halo orbit.

\textbf{Cost function.} An important parameter that represents the cost of a space mission is the variation of velocity $\Delta V$ required in the maneuvers, as shown in Equation \ref{J(y)}.
\begin{equation}
J(\textit{\textbf{y}}) = \mid \Delta \textit{\textbf{V}} \mid. 
\label{J(y)}
\end{equation}

\textbf{Constraints.} The intersection in the configuration space between the initial parking orbit around the Moon and the transfer trajectory provided by the stable invariant manifold of the final halo orbit around $L_2$ represents the most important constraint. Thus, the constraint set is represented by the equality $(c_{eq}=0)$, with 
\begin{equation}
c_{eq} = \begin{bmatrix}
\textbf{r}_t(t_{IM})- \textbf{r}_{IM}(t_{IM})\\ h_p-600 \\ h_a - 20,000
\end{bmatrix},
\label{ceq}
\end{equation}
where $\textbf{r}_t$ is the position of the lunar parking orbit that intersects the transfer trajectory at point $\textbf{r}_{IM}$ at the instant $t_{IM}$. Furthermore, $h_p$ and $h_a$ are the periselene and the aposelene altitude, respectively. Mathematically, we can write the transfer problem using just one optimal impulsive maneuver as a restricted minimization, prescribed as
\begin{equation}
\underset{\textit{\textbf{y}}}{min}~J(\textit{\textbf{y}}) ~~ \mbox{such that} ~~ c_{eq} = 0.
\label{c}
\end{equation}

Given this approach, all the three trajectories involved in this transfer maneuver are varied to provide the minimun cost solution, namely, the lunar parking orbit, the manifold-guided transfer trajectory, and the halo orbit around $L_2$. To find the candidates for the optimal solutions, the nonlinear optimization routine (fmincon) developed by Matlab company was employed. An outline of the transfer design logic can be seen in Table \ref{tab:11} \citep{Cipriano}.
\begin{table}
	\caption{Transfer Project Algorithm.}
	\label{tab:11}       
	\begin{tabular}{l}
		
		\textbf{procedure} INITIALIZATION   \\
		~ Set the CRTBP as default dynamical model  \\
		~ Select $C_j$ of target halo orbit \\
		~ Select manifold branch flying toward the Moon \\
		\textbf{procedure} end procedure  \\
		\\
	    \textbf{procedure} MANIFOLD SCAN FOR INITIAL GUESS\\
	    ~~ Set bounds for the time along the target halo, $t_{1}$ $\in$ [0, T]\\
        ~~ Set maximum time along the stable manifold, $t^{(max)}_{2}$\\
        ~~ Discretize $t_{1}$ by $dt_{1}$ to get $n_t$ discrete values\\
		~~ Initialize vector $\tau$ $\in$ $\mathbb{R}^{n_t\times 3}$\\
		~~ \textbf{for} $t_{1}$ = 0 $\rightarrow$ T by $dt_{po}$ with index $k$ \textbf{do}\\
        ~~~~ Get stable manifold state, $\textit{\textbf{x}}_{s}$, for current $t_{1}$ and $t_2^{(max)}$\\
        ~~~~ See Equation \ref{unst}\\
        ~~~~ Find $t_{2}$ at which altitude, $h_p$, is closest to 200 km\\
        ~~~~ Store ($t_{1}$, $t_{2}$, $h_p$) in the $k$-th row of vector $\tau$\\
        ~~ \textbf{end for}\\
        \textbf{end procedure}\\
        \\
        \textbf{procedure} TRANSFER MANEUVER\\
        ~~ Initialize vector $\mathbf{\Gamma}$ $\in$ $\mathbb{R}^{n_t\times 16}$\\
        ~~ \textbf{loop} in $\mathbf{\tau}$ with index $j$\\
        ~~~~ \textbf{repeat}\\
        ~~~~~~ Randomly initialize injection orbit elements in\\
        ~~~~~~ $e$ = ($h_a$, $i$, $\Omega$, $\omega$, $\theta$)\\
        ~~~~~~ Solve for $\Delta V_{IM}$ using $\textbf{e}$ and j-th row of $\tau$ as first guess \\
        ~~~~~~ Equation \ref{c}\\
        ~~~~ \textbf{until} convergence is attained\\
        ~~~~ Store optimization results, \\
        ~~~~ ($\Delta V_{IM}$, $\mathbf{y}^{(opt)}$), in $\mathbf{\Gamma}$ j-th row\\
        ~~ \textbf{end loop}\\
\textbf{end procedure}

	\end{tabular}
\end{table}

\section{Numerical Results}

In this section, we present results obtained for two cases. In subsection \ref{otimalsolution1}, both values of  $h_p$ and $h_a$ were kept fixed, namely, $h_ p$ = 200 km and $h_ a$ = 20,000 km; while in subsection \ref{otimalsolution2},  the results were obtained considering only the value of $h_a$ fixed, for $h_ a$ = 20,000 km.

The adopted algorithm requires an initial guess that is iteratively improved by the  optimization process towards an optimal solution. As a result, choosing the initial guess is crucial in determining  how quickly the algorithm converges. Moreover, often the studied function has multiple extremes, and it is necessary to use different initial guesses to find the local and global extremes. Knowing this, several initial guesses were investigated in order to obtain a representative set of candidates for the optimal solutions. 

To determine the geometric configuration of the parking orbit, as shown in Table \ref{tab:1}, we used as initial guess in the numerical simulation the fixed values of $ h_p $ = 600 km, $ h_a $ = 20,000 km  (therefore  values of $a$ and $e$ are fixed), $ \omega $ = 240\textdegree and $\theta$ = 0\textdegree.
By the other side, the variables  varied in the initial guess set were the inclination and right ascension of the ascending node (RAAN).

We initially used zero inclination and we vary the inclination up to 180\textdegree in the step of 5 degrees. We also varied the RAAN from 0\textdegree to 360\textdegree in the step of 10\textdegree. 
\begin{table}
	\caption{Keplerian elements used as initial guess of the optimization algorithm.}
	\label{tab:1}       
	\begin{tabular}{lll}
		
		Initial inclination (degree)  & i $\in$ [0, 180\textdegree] \\
		Initial RAAN (degree) & $\Omega$ $\in$ [0, 360\textdegree] \\
		initial semimajor axis (km) & a = 12037.4 \\
		initial eccentricity (adim) & e = 0.8058 \\
		Initial periselene anomaly (degree)  & $\omega$ = 240 \\
		Initial true anomaly (degree)  & $\theta$ = 0 \\
		
	\end{tabular}
\end{table}

\subsection{Optimal solutions considering  $h_p$ = 200 km and $h_a$ = 20,000 km }
\label{otimalsolution1}

\begin{figure}[!htbp]
	\centering\includegraphics[scale=0.37]{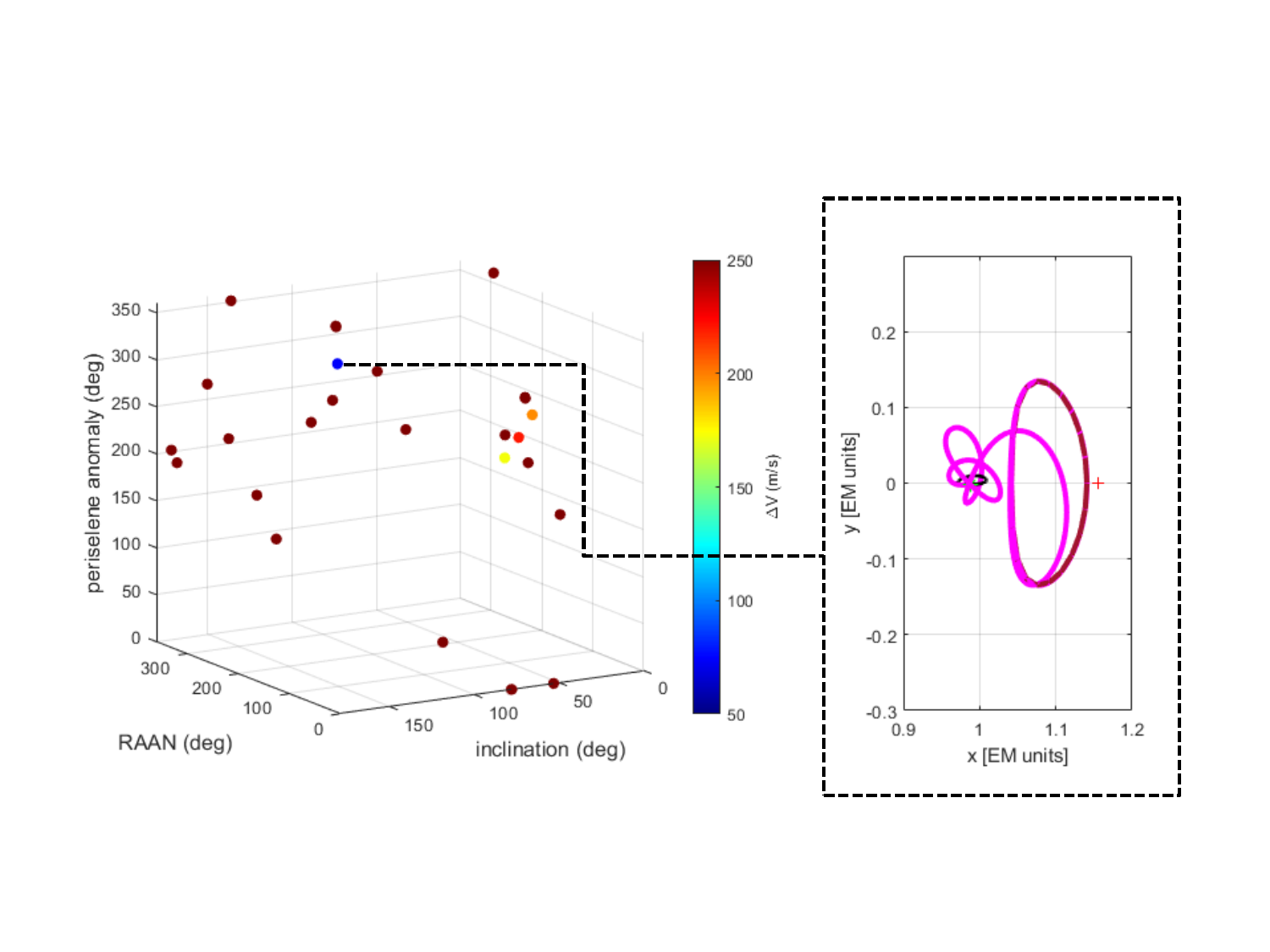}
	\caption{Candidates for optimal solutions obtained for fixed Jacobi constant $C_j$ = 3.07, presenting the Keplerian elements of the correspondent lunar parking orbit (left frame) and the full trajectory of the best solution for this $C_j$ value (right frame). Color code are detailed in the text. Characteristics of the Kleperian elements and the transfer trajectory, for $Cj$ = 3.07, are shown in Tables \ref{table2} and \ref{table3}, respectively.}
	\label{CJ307CPMoptimal}
\end{figure}

\begin{figure}[!htbp]
	\centering\includegraphics[scale=0.37]{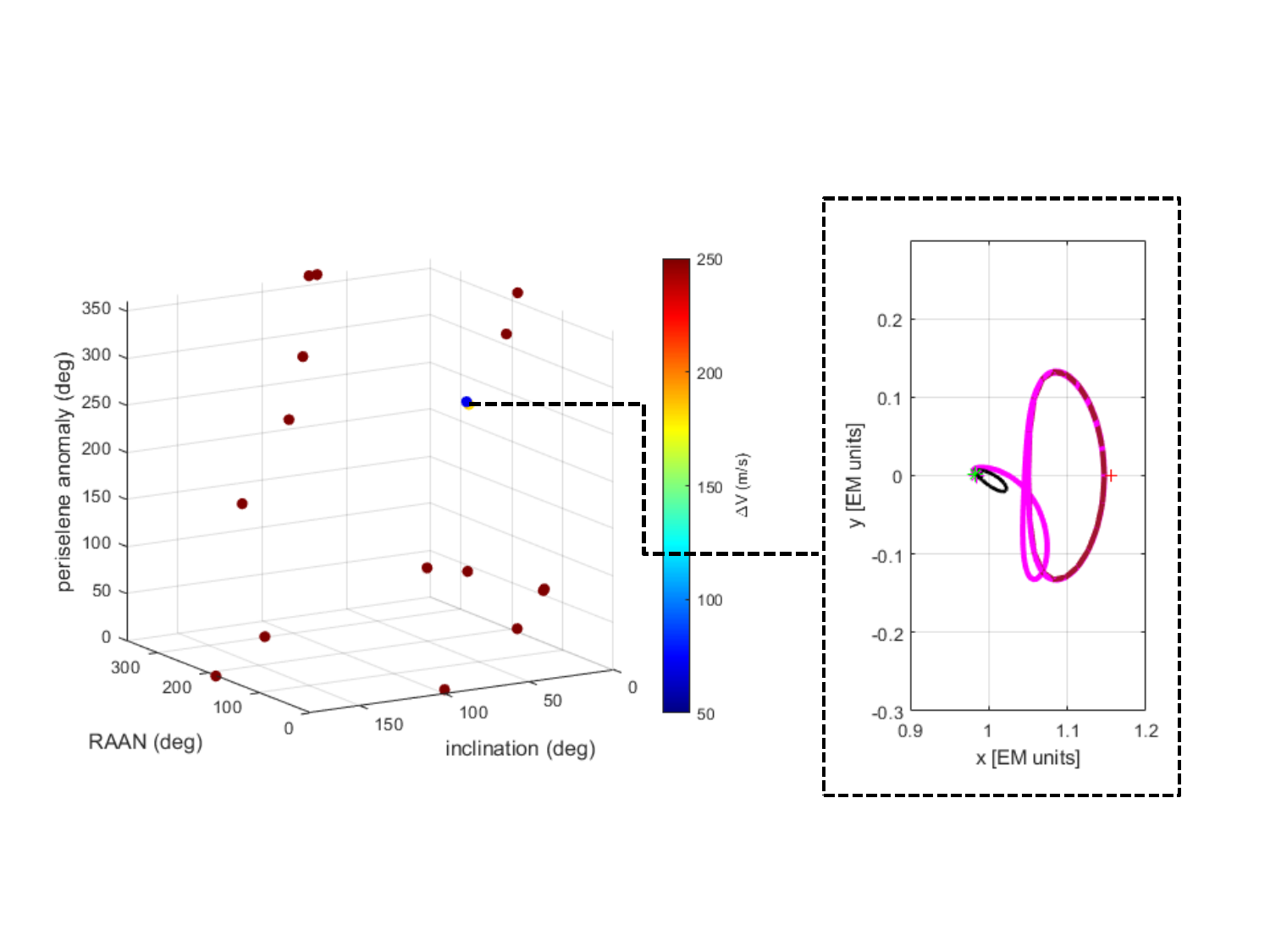}
	\caption{Candidates for optimal solutions obtained for fixed Jacobi constant $C_j$ = 3.08, presenting the Keplerian elements of the correspondent lunar parking orbit (left frame) and the full trajectory of the best solution for this $C_j$ value (right frame). Color code are detailed in the text. Characteristics of the Kleperian elements and the transfer trajectory, for $Cj$ = 3.08, are shown in Tables \ref{table2} and \ref{table3}, respectively.}
	\label{CJ308general}
\end{figure}

\begin{figure}[!htbp]
	\centering\includegraphics[scale=0.37]{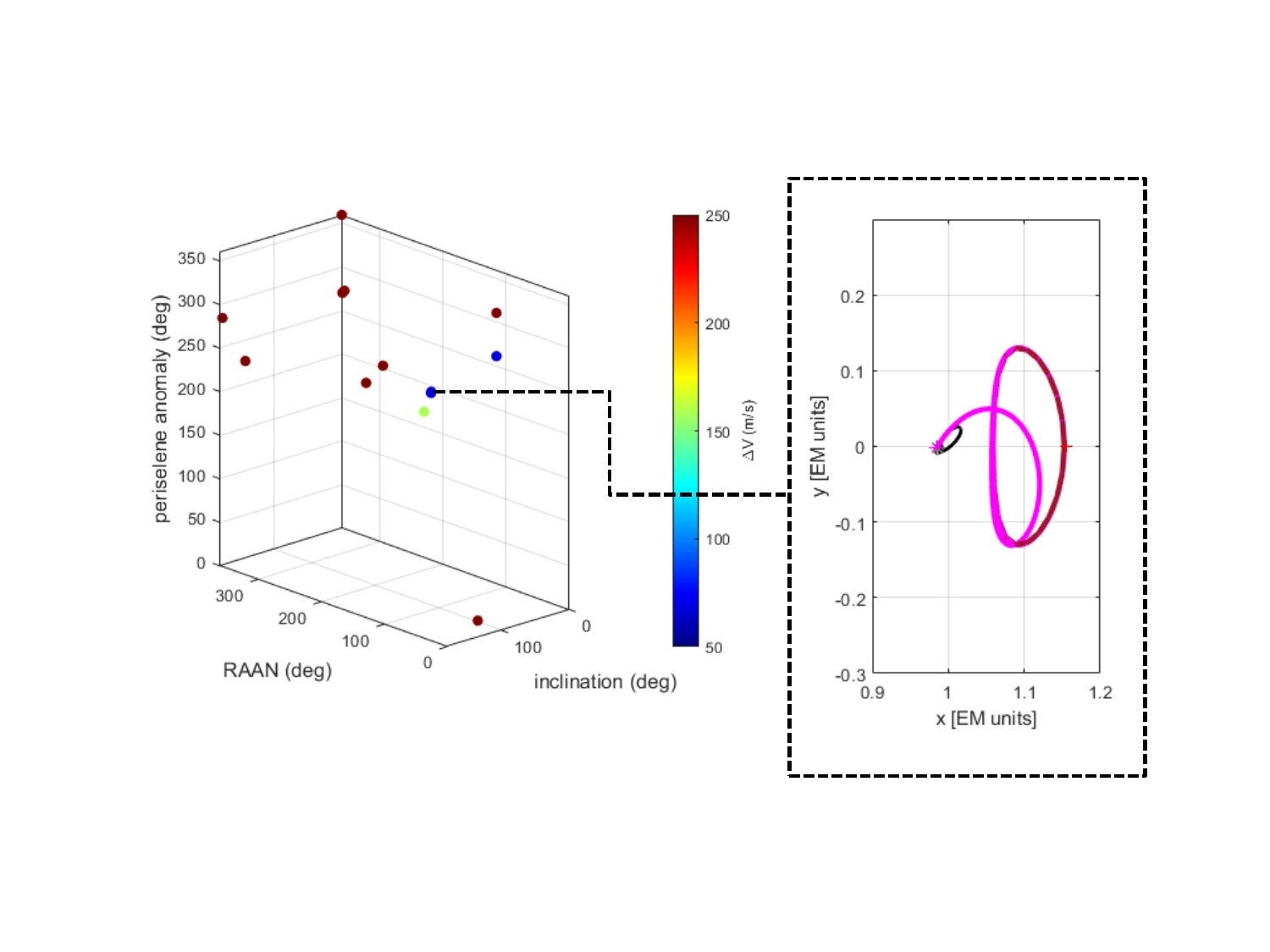}
	\caption{Candidates for optimal solutions obtained for fixed Jacobi constant $C_j$ = 3.09, presenting the Keplerian elements of the correspondent lunar parking orbit (left frame) and the full trajectory of the best solution for this $C_j$ value (right frame). Color code are detailed in the text. Characteristics of the Kleperian elements and the transfer trajectory, for $Cj$ = 3.09, are shown in Tables \ref{table2} and \ref{table3}, respectively.}
	\label{generalC9}
\end{figure}

\begin{figure}[!htbp]
	\centering\includegraphics[scale=0.37]{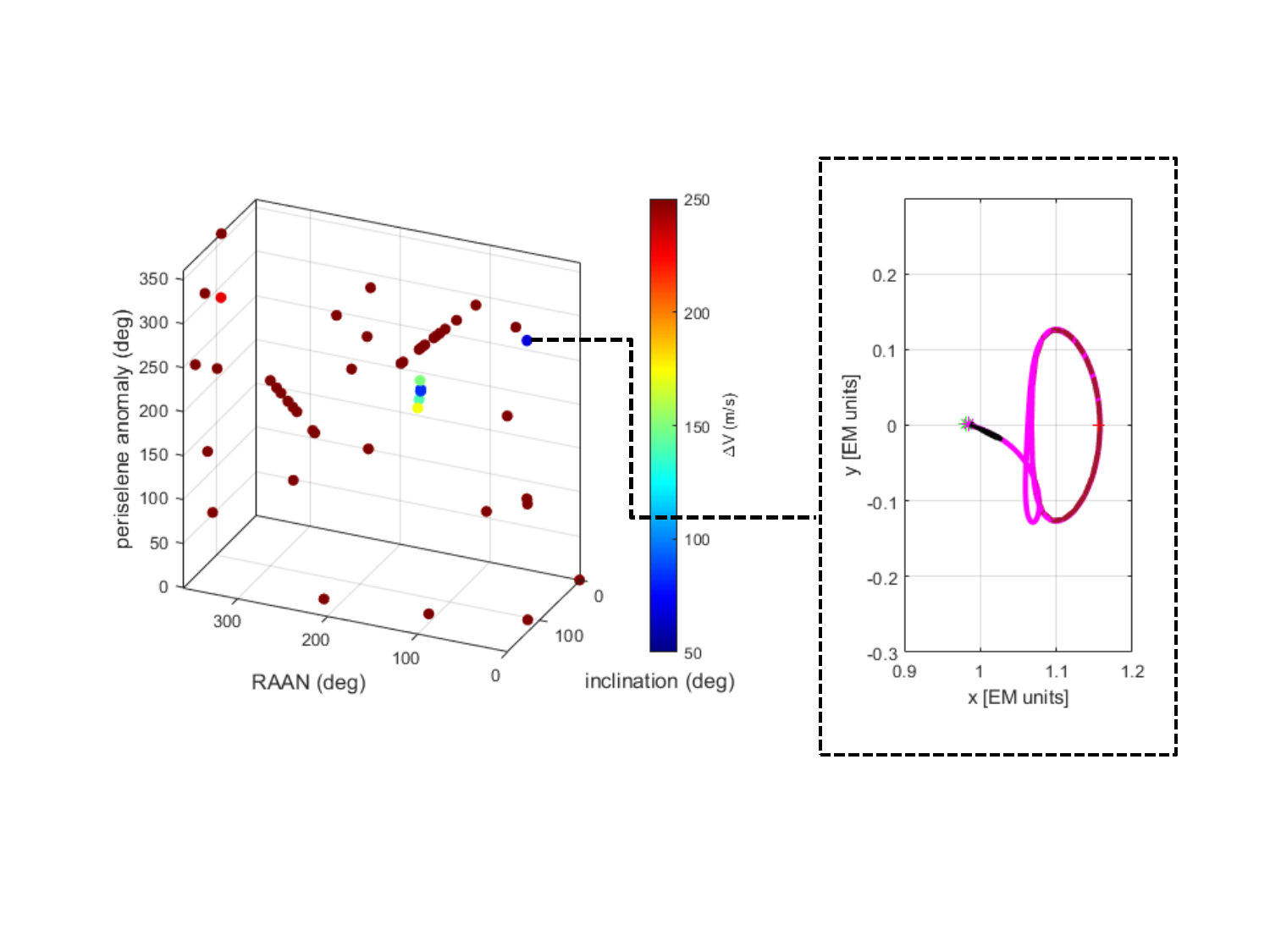}
	\caption{Candidates for optimal solutions obtained for fixed Jacobi constant $C_j$ = 3.10, presenting the Keplerian elements of the correspondent lunar parking orbit (left frame) and the full trajectory of the best solution for this $C_j$ value (right frame). Color code are detailed in the text. Characteristics of the Kleperian elements and the transfer trajectory, for $Cj$ = 3.10, are shown in Tables \ref{table2} and \ref{table3}, respectively.}
	\label{CJ310general}
\end{figure}

\begin{figure}[!htbp]
	\centering\includegraphics[scale=0.36]{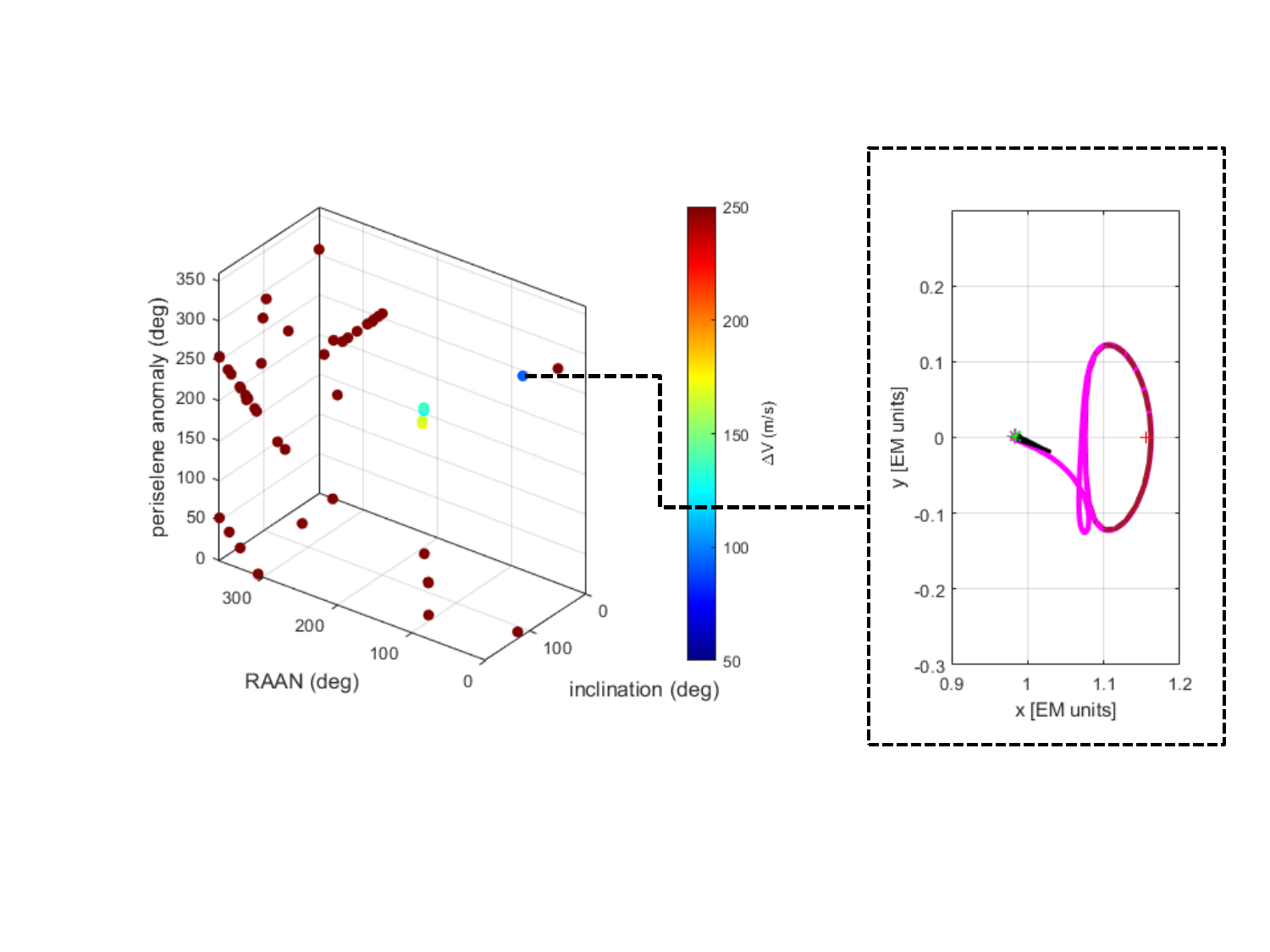}
	\caption{Candidates for optimal solutions obtained for fixed Jacobi constant $C_j$ = 3.11, presenting the Keplerian elements of the correspondent lunar parking orbit (left frame) and the full trajectory of the best solution for this $C_j$ value (right frame). Color code are detailed in the text. Characteristics of the Kleperian elements and the transfer trajectory, for $Cj$ = 3.11, are shown in Tables \ref{table2} and \ref{table3}, respectively.}
	\label{CJ311general}
\end{figure}

Several optimal solutions were found for transfer orbits with different values of the Jacobi constant, ranging from $Cj = 3.07$ to $Cj = 3.11$ as can be seen in Figures \ref{CJ307CPMoptimal} to \ref{CJ311general}.

The left frames of Figs. \ref{CJ307CPMoptimal} to \ref{CJ311general} present the Keplerian elements of the lunar parking orbits of the solution provided by the algorithm.
The color code indicates the velocity variation ($\Delta V$) required for each of these transfer solutions. It is remarkable that, at least, two solutions for each Jacobi constant level present $\Delta V$ within the range of values sought in this investigation, namely, $\Delta V$ $\leq$ 200 m/s. 
We remind that, in this subsection, the other two Keplerian elements of the lunar parking orbit, the semi-major axis and eccentricity, are fixed constant (as shown in Table \ref{tab:1}). In all the solutions obtained as candidates for the optimal solutions, we observe that the maneuver for insertion in the stable manifold of the halo orbit is performed at the perilune of the parking orbit, that is, the true anomaly is approximately zero in all the cases. The number of iterations performed to find the solutions varies depending on the initial guess. If the initial estimate is far from the solution sought, the number of interactions can be 150. On the other hand, if the initial estimate is close to the solution, the number of interactions is around 15.

The right frame of Figs. \ref{CJ307CPMoptimal} to \ref{CJ311general} shows the $xy$ projection of the full trajectory of the best candidate for the optimal transfer solution obtained for each investigated Jacobi Constant level, from $Cj = 3.07$ to $Cj = 3.11$. 
The black curve represents the parking orbit around the Moon of these optimal candidates, while the pink and the blue curves  depict the transfer leg in the stable manifold of the halo orbit around $L_2$, and the target halo orbit itself.

Note that all the optimal solution candidates contain a prograde orbit as the lunar parking orbit ($i~ < ~90^\circ$).

Tables \ref{table2} and \ref{table3} provide the main features of the best solution obtained in Figs. \ref{CJ307CPMoptimal} to \ref{CJ311general}, respectively.

\begin{table*}
\centering
\caption{Kleperian elements of the lunar parking orbit of the solutions presented in the right frame of  Figures \ref{CJ307CPMoptimal} to \ref{CJ311general}, i.e., the  best solution found for each Jacobi constant considered from 3.07 to 3.11.}
\label{table2}
\begin{tabular}{lllllll}
		$a$ [km]  & $e$ [adim]& i [deg] &$\Omega$ [deg]& $\omega$ [deg] & $\theta$ [deg]& $Cj$ [adim] \\
			12037.45 & 0.8058 &74.7500& 353.7580 & 270.122 & 0& 3.07\\		
			12037.45 & 0.8058 &76.0362& 36.2577 & 296.360 & 0& 3.08\\
			12037.45 & 0.8058 &85.3900& 126.3270 & 237.586 & 0& 3.09\\
			12037.45 & 0.8058 &88.3273& 19.0030 & 308.703 & 0& 3.10\\
			12037.45 & 0.8058 &95.3300& 13.07569 & 313.076 & 0& 3.11\\
	\end{tabular}
	\end{table*}

\begin{table}
\centering
\caption{Other relevant parameters of the best solution obtained for each $C_j$ value (as shown in the right frame of Figures \ref{CJ307CPMoptimal} to \ref{CJ311general}, respectively). These values are related to the lunar parking orbit ($t_1$), the transfer phase ($t_2$) and the correspondent halo orbit ($T$), besides the cost of the single maneuver  ($\Delta V$ ).}
\label{table3}
\begin{tabular}{lllll}
		 $t_{1}$ [adim]  & $t_{2}$ [adim]& T [days] &$\Delta V$ [m/s]& $Cj$ [adim] \\
			0.03048 & 11.32095 &37.361& 72.247 & 3.07\\		
			3.0527 & 8.3347 &36.193& 70.4370 & 3.08\\
			1.42418 & 7.2891 &31.652& 67.939 & 3.09\\
			1.0387 & 7.2454 &31.163& 66.798 & 3.10\\
			1.3114 & 6.9967 &30.3833& 74.026  & 3.11\\
\end{tabular}
\end{table}
 
	
		
\begin{figure}[!htbp]
	\centering\includegraphics[scale=0.6]{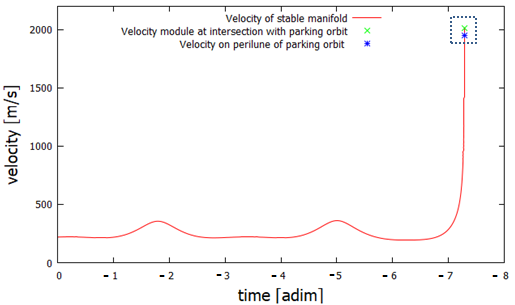}
	\caption{Stable manifold velocity module as a function of time.}
	\label{velocity}
\end{figure}
\begin{figure}[!htbp]
	\centering\includegraphics[scale=0.6]{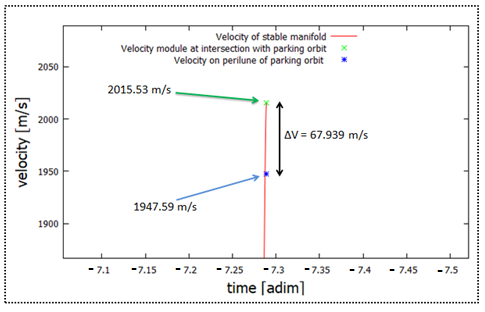}
	\caption{Magnification of Figure \ref{velocity} around velocity near the Lunar surface.}
	\label{velocityzoom}
\end{figure}

Figures \ref{velocity} and \ref{velocityzoom} examine the lowest cost solution obtained for $C_j$ = 3.09, presenting the retrograde time evolution of the velocity  module of the spacecraft along the transfer orbit from the halo orbit departure (at t = 0) until  the vicinity of the Moon is reached (t $\approx$ -7 adimensional time unit). We observe that near the halo orbit, velocity module values are low, but as the spacecraft approaches the Moon, the potential energy absolute value increases causing a very fast growth of the velocity module. The magnification presented in Fig. \ref{velocityzoom} shows the relevant values for the maneuver cost computation, i.e., the transfer trajectory velocity at the maneuver point (green x), 2015.53 m/s; the modulus of the parking orbit velocity in the perilune (blue asterisk), 1947.59 m/s; and the required $\Delta V$= 67.939 m/s at the patching orbit. Note that Fig. \ref{velocityzoom} is a zoom of the region delimited by the blue dashed square in Fig. \ref{velocity}. These observations explain why in all the best candidates for optimal solutions obtained for each Jacoby constant, the maneuver is performed at the perilune of the lunar parking orbit (the highest velocity point of the parking orbit). We also remark that the used constraints in this subsection, i.e.,  $h_p$= 600 km and $h_a= 20,000$ km imply that $V_p$ = 1947.59 m/s and $V_a$ = 207 m/s.

\begin{figure}[!htbp]
	\centering\includegraphics[scale=0.6]{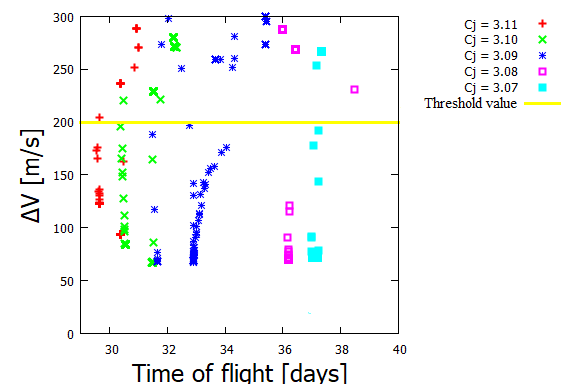}
	\caption{Time of flight versus the total  $\Delta V$ cost of the candidates for optimal solutions.}
	\label{draw_3}
\end{figure}

Figure \ref{draw_3} shows the time of flight of the transfer phase $t_2$ versus the required $\Delta V$ of the candidates for the best solutions obtained for different Jacobi constant. The horizontal yellow line indicates the  threshold value  of our investigation for reference ($\Delta V$ = 200 m/s). 

The solutions are grouped according to the Jacobi constant in five sets: red crosses for $Cj$ = 3.11, green x for $Cj$ = 3.10, blue asterisk for $Cj$ = 3.09, purple square for $Cj$ = 3.08, and cyan filled squares for $Cj$ = 3.07. We observe, from Figure \ref{draw_3}, that, as we increase the energy of the system, that is, when we decrease the value of the Jacobi constant, the time along of manifold becomes larger. This occurs due to the fact that the increase in energy makes the halo orbit larger, taking more time for the spacecraft to reach the final destination. 

Note in Figure \ref{draw_3} (and Table \ref{table3}) that, for lower time of flight, the value of the optimal $\Delta V$ decreases as we increase the energy (decrease the Jacobi constant) until reaching a minimum when $Cj$ = 3.10. After that, optimal solutions $\Delta V$ tend to grow as the value of the Jacobi constant decreases even more. Additional numerical experiments were performed for $C_j$ ranging from 3.02 to 3.16. 
However, for $C_j<3.07$ and for $C_j>3.11$, results show that the required $\Delta V$ of the solutions obtained by the optimization algorithm increases gradually.

The data shown in Figure \ref{draw_3} are important for planning a space mission. We know that, in a real mission, it is impossible for the Cubesat (or spacecraft) to follow the nominal trajectory, due to factors that were not simulated, and numerical errors. But we can approximate the real orbit of the nominal orbit as much as possible, so that, the Jacobi constant, although not exactly 3.09, will be a value around this number, thus maintaining the minimum $\Delta V$. 

We can understand why the transfer cost has a well-defined minimum range. This is due to the fact that with high energy, i.e. low Jacobi constant, the stable manifold passes far from the Moon, so it is necessary to spend a high fuel consumption to perform the IM \citep{Dio}. On the other hand, when we have low energy (Jacobi constant high), the stable manifolds of the halo orbits around $L_2$ pass too close to the Lunar surface \citep{Dio}. This makes the velocity of the stable manifold higher compared to the velocity of the parking orbit, and thus a high $\Delta V$ is required to insert the spacecraft into the stable manifold from the parking orbit.

\subsection{Optimal solutions considering  hp varying and ha fixed, where ha = 20,000 km }
\label{otimalsolution2}

\begin{figure}[!htbp]
	\centering\includegraphics[scale=0.6]{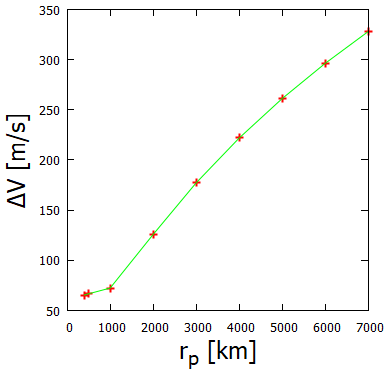}
	\caption{Total fuel cost for inserting the spacecraft into a stable manifold trajectory as a function of periselene altitude.}
	\label{rpxV}
\end{figure}

Using the Jacobi constant $Cj$ = 3.09, we search for optimal solutions of $\Delta V$ as a function of periselene altitude $h_p$. In Figure \ref{rpxV} we plot the lowest cost transfers obtained as a function of  the periselene altitude $h_p$.
We see from Figure \ref{rpxV} that as the periselene altitude becomes smaller, the total $\Delta V$ to perform the maneuver also decreases. For these results, we maintain the aposelene altitude constant ($h_a$ = 20,000 km) and we vary only $h_p$. Since the semi major-axis and orbit eccentricity depend on $r_p$ and $r_a$, the results show that as we increase the eccentricity or decrease the semi major-axis of the parking orbit, the total fuel expenditure to perform the maneuver becomes smaller. The best solutions, that is, the results with the smallest $\Delta V$ to perform the transfer occur approximately in the periselene of the orbit. 
This is due to the fact that the lower the periselene altitude, the faster the spacecraft moves in this position, decreasing the $\Delta V$ required to perform the transfer and leave the Hill region around the Moon.

\section{Conclusion}

In this paper, the optimization of a full space mission trajectory  comprising a lunar parking orbit, a manifold-guided transfer, and a final halo orbit around the $L_2$ point of the  Earth-Moon system was investigated. Our main goal is to look for solutions of minimum cost, considering the application of a single impulsive maneuver. Several candidates for the optimal solutions were found using different Jacobi constant values. Depending on the Jacobi constant used, it was possible to find transfers using $ \Delta V \leq $ 70 m/s.
We also determined the time of flight of the transfer between the initial parking orbit (around the Moon) to halo orbit (around $L_2$). With the optimal solutions found, the flight time along of manifold guided transfer lasts from 30 to 40 days, depending on the Jacobi constant used. Also we observe that, as the value of the Jacobi constant increases, the time required to reach halo orbit around the equilibrium point $L_2$ decreases. Numerical evidence shows that for $Cj$ = 3.09, $\Delta V$ fuel consumption reaches a global minimum with respect to the Jacobi constant values used in this study. 
Finally, we find optimal solutions of $\Delta V$ as a function of perilune radius of the parking orbit. It was possible to verify that the smaller the perilune of the orbit, the lower the fuel consumption required to insert the Cubesat into the transfer trajectory and reach the halo orbit. 

\section{Acknowledgments}
The authors wish to express their appreciation for the support provided by grants\# 422282/2018-9, 150678/2019-3, 309089/2021-2 from the National Council for Scientific and Technological Development (CNPq), grants\# 2016/24561-0, 2016/18418-0, 2016/24970-7, 2018/07377-6 from S\~ao Paulo Research Foundation (FAPESP) and grants\# 0000-000 from  RUDN University. We also thank the the financial support from the National Council for the Improvement of Higher Education (CAPES). This paper has been supported by the RUDN University Strategic Academic Leadership Program. A part of this work was pursued at the Politecnico di Milano, Department of Aerospace Science and Technology.

\end{document}